\documentclass[traditabstract,letter]{aa} 

\usepackage{graphicx}
\usepackage{txfonts}
\newcommand{\vsini}{\mbox{$v \sin i$}}

\newcommand{\etal}{et\,al.}
\newcommand{\kmps}{km\,s$^{-1}$}

\def\qps{$Q^{\prime}_{\rm *}$}

\begin{document}

\title{WASP-43b: The closest-orbiting hot Jupiter}
\titlerunning{WASP-43b}
\authorrunning{Hellier et al.}

\author{
Coel Hellier\inst{1}, 
D.R. Anderson\inst{1}, 
A. Collier Cameron\inst{2}, 
M. Gillon\inst{3}, 
E. Jehin\inst{3}, 
M. Lendl\inst{4}, 
P.F.L. Maxted\inst{1}, 
F. Pepe\inst{4}, \\
D. Pollacco\inst{5}, 
D. Queloz\inst{4}, 
D. S\'egransan\inst{4}, 
B. Smalley\inst{1}, 
A.M.S. Smith\inst{1}, 
J. Southworth\inst{1}, 
A.H.M.J. Triaud\inst{4}, 
S. Udry\inst{4} \\  \&\ 
R.G. West\inst{6}}

\institute{Astrophysics Group, Keele University, Staffordshire, ST5 5BG, UK
\and SUPA, School of Physics and Astronomy, University of St.\ Andrews, North Haugh,  Fife, KY16 9SS, UK
\and Institut d'Astrophysique et de G\'eophysique, Universit\'e de
Li\`ege, All\'ee du 6 Ao\^ut, 17, Bat. B5C, Li\`ege 1, Belgium
\and Observatoire astronomique de l'Universit\'e de Gen\`eve
51 ch. des Maillettes, 1290 Sauverny, Switzerland
\and Astrophysics Research Centre, School of Mathematics \& Physics, Queen's University, University Road, Belfast, BT7 1NN, UK
\and Department of Physics and Astronomy, University of Leicester, Leicester, LE1 7RH, UK}

\abstract
{We report the discovery of WASP-43b, a hot Jupiter 
transiting a K7V star every 0.81 d. At 0.6-M$_{\odot}$ 
the host star has the lowest mass of any star hosting a hot Jupiter.  It also shows a 15.6-d rotation period.   
The planet has
a mass of 1.8\,M$_{\rm Jup}$, a radius of 
0.9 R$_{\rm Jup}$, and with a semi-major axis of only 0.014 AU
has the smallest orbital distance of any known hot Jupiter.
The discovery of such
a planet around a K7V star shows that planets with apparently
small remaining lifetimes owing to tidal decay of the orbit are
also found around stars with deep convection zones.}

\keywords{stars: individual (WASP-43) --- planetary systems}

\maketitle

\section{Introduction}
As planet discoveries increase we begin to see patterns
in their distribution, and to find the rarer systems that
mark the edges of the envelope. The ground-based 
transit searches such as WASP (Pollacco \etal\ 2006) and HAT
(Bakos \etal\ 2002) are particularly suitable for finding the
systems that delineate the cut-off of hot Jupiters 
as orbital radius decreases.   This distribution is expected
to tell us about several processes, including disk migration
and possible `stopping mechanisms' (e.g.\ Matsumura, Pudritz \&\ 
Thommes 2007), 
third-body processes, such as scattering and the Kozai mechanism,
that can move planets onto eccentric orbits that circularize
at short periods (e.g.\ Guillochon \etal\ 2010), and the 
effect of tidal interactions with the host star (e.g.\ Matsumura,
Peale \&\ Rasio 2010). 

The WASP-South camera array has
been monitoring stars of magnitude 9--13 since 2006, and, 
in conjunction with radial-velocities
from the Euler/CORALIE spectrograph, is now responsible for
the majority of transiting hot Jupiters currently known in the 
Southern hemisphere (see Hellier \etal\ 2011a).  Here we report 
the discovery of WASP-43b, which has the smallest semi-major
axis of any known hot Jupiter.

\begin{figure}
\hspace*{-5mm}\includegraphics[width=9cm]{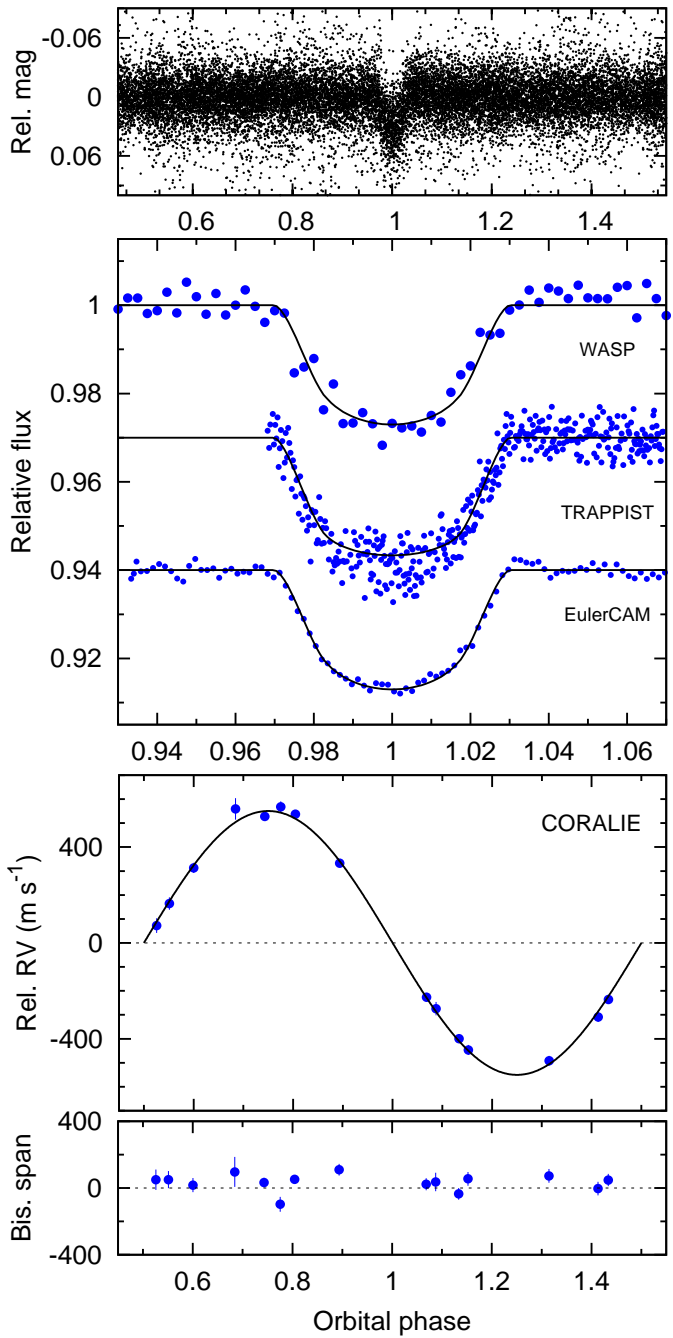}\\ [-2mm]
\caption{(Top) The WASP-South lightcurve folded on the 0.81-d transit
period. (Second panel) The binned WASP data with (offset) the
TRAPPIST ($I$+$z$) and Euler (Gunn r) transit lightcurves,
with the fitted MCMC model.  
(Third) The CORALIE radial
velocities with the fitted model.
(Lowest) The bisector spans; the absence of any correlation with
radial velocity is a check against transit mimics (Queloz et\,al.\
2001).}
\end{figure}

\begin{table}
\caption{CORALIE radial velocities of WASP-43.\protect\rule[-1.5mm]{0mm}{2mm}} 
\label{rv-data} 
\begin{tabular*}{0.5\textwidth}{@{\extracolsep{\fill}}cccr} 
\hline 
BJD\,--\,2400\,000 & RV & $\sigma$$_{\rm RV}$ & Bisector\\ 
 & (km s$^{-1}$) & (km s$^{-1}$) & (km s$^{-1}$)\\ [0.5mm]
\hline
55205.7594 & --3.058 & 0.013 & \rule{0mm}{5mm}0.052\\
55325.6232 & --4.041 & 0.021 & ~0.055\\
55327.5745 & --3.430 & 0.026 & ~0.050\\
55328.5441 & --3.067 & 0.014 & ~0.033\\
55334.5030 & --3.821 & 0.018 & ~0.023\\
55359.4824 & --3.026 & 0.022 & --0.098\\
55362.5333 & --3.522 & 0.031 & ~0.050\\
55364.4596 & --3.262 & 0.017 & ~0.110\\
55375.4741 & --3.830 & 0.018 & ~0.048\\
55376.4911 & --3.036 & 0.045 & ~0.097\\
55378.4837 & --3.994 & 0.018 & --0.035\\
55379.5246 & --3.904 & 0.021 & --0.003\\
55380.4904 & --3.282 & 0.021 & ~0.017\\
55391.4617 & --3.869 & 0.028 & ~0.036\\
55392.4602 & --4.086 & 0.021 & ~0.072\\
\hline
\multicolumn{4}{l}{Bisector errors are twice RV errors} 
\end{tabular*} 
\end{table} 

\begin{figure}
\hspace*{-5mm}\includegraphics[width=9cm]{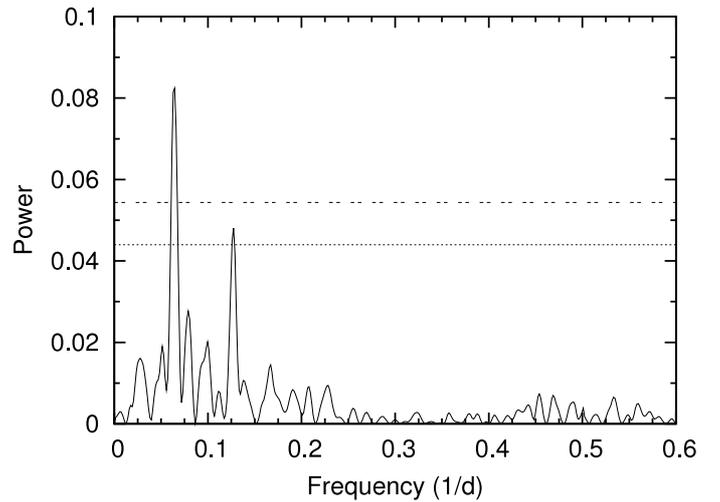}\\ [-2mm]
\caption{A periodigram of the 2009 WASP-South photometry, showing a
15.6-d modulation.  The horizontal lines show the 0.01 and
0.001 false-alarm probability levels.}  
\end{figure}

\section{Observations}
The WASP project uses 8-camera arrays that cover
450 square degrees of sky with a typical 
cadence of 8 mins.
The WASP surveys are described in \cite{sw}
while a discussion of our planet-hunting methods can 
be found in Collier-Cameron \etal\ (2007a) and Pollacco \etal\ (2007).

WASP-43 is a $V$ = 12.4, K7V star in the
constellation Sextans.  It was flagged as a planet
candidate based on WASP-South data obtained during 2009 January--May,
and has been further observed by both WASP-South and SuperWASP-North
over 2010 January--May, leading to a total of 13\,768 data points.  
A putative 0.81-d transit period led to radial-velocity
followup with the CORALIE spectrograph on the Euler 1.2-m telescope
at La Silla. Fourteen radial-velocity measurements over  
2010 January--July (Table~1) showed that the 
transiting body is a 1.8-M$_{\rm Jup}$ planet.  On 2010 December 07 
we obtained a transit lightcurve with the TRAPPIST 0.6-m telescope
in a passband of $I$+$z$, while on 2010 December 29 we obtained
a further transit lightcurve with EulerCAM in a Gunn $r$  
passband (Fig.~1).

\begin{table} 
\caption{System parameters for WASP-43.\protect\rule[-1.5mm]{0mm}{2mm}}  
\begin{tabular}{lclc}
%\begin{tabular}{lc}%{0.5\textwidth}{@{\extracolsep{\fill}}lc} 
\hline
%Parameter (Unit) & Value \rule{0mm}{5mm} \\ [0.5mm] 
%\hline
\multicolumn{4}{l}{Stellar parameters from spectroscopic analysis.\rule[-1.5mm]{0mm}{2mm}} \\
\hline 
%\\ \hline
\multicolumn{4}{l}{RA\,=\,10$^{\rm h}$19$^{\rm m}$38.01$^{\rm s}$, 
Dec\,=\,--09$^{\circ}$48$^{'}$22.5$^{''}$ (J2000)\rule{0mm}{5mm}}\\
$V$ mag & 12.4  &
Spectral type & K7V \\
$T_{\rm eff}$ (K)      & 4400 $\pm$ 200  &
$\log g$      & 4.5 $\pm$ 0.2 \\
$\xi_{\rm t}$ (km\,s$^{-1}$)    & 0.5 $\pm$ 0.3 & 
$v\,\sin i$ (km\,s$^{-1}$)     & 4.0 $\pm$ 0.4 \\
{[Fe/H]}   &$-$0.05 $\pm$ 0.17 &
{[Na/H]}   &   0.57 $\pm$ 0.18 \\
{[Al/H]}   &   0.30 $\pm$ 0.12 &
{[Si/H]}   &   0.01 $\pm$ 0.20 \\
{[Ca/H]}   &   0.19 $\pm$ 0.21 & 
{[Sc/H]}   &   0.06 $\pm$ 0.16 \\ 
{[Ti/H]}   &   0.20 $\pm$ 0.21 &
{[V/H]}    &   0.42 $\pm$ 0.25 \\ 
{[Cr/H]}   &$-$0.03 $\pm$ 0.21 & 
{[Mn/H]}   &   0.36 $\pm$ 0.18 \\ 
{[Co/H]}   &   0.24 $\pm$ 0.05 & 
{[Ni/H]}   &   0.08 $\pm$ 0.20 \\
log A(Li)  &   $<$ 0.2 $\pm$ 0.3 &
Distance   &   80 $\pm$ 20 pc \\ [0.8mm] \hline
\end{tabular}
\begin{tabular}{lc}
\multicolumn{2}{l}{Parameters from MCMC analysis.\rule[-1.5mm]{0mm}{5mm}} \\
\hline 

$P$ (d) & 0.813475 $\pm$ 0.000001\\
$T_{\rm c}$ (HJD) & 2455528.86774 $\pm$ 0.00014\\
$T_{\rm 14}$ (d) & 0.0483 $\pm$0.0011\\
$T_{\rm 12}=T_{\rm 34}$ (d) & 0.0110$\pm$ 0.0017\\
$\Delta F=R_{\rm P}^{2}$/R$_{*}^{2}$ & 0.0255 $\pm$ 0.0012\\
$b$ & 0.66$^{+ 0.04}_{- 0.07}$\\
$i$ ($^\circ$)  & 82.6$^{+ 1.3}_{- 0.9}$\\
$K_{\rm 1}$ (m s$^{-1}$) & 550.3 $\pm$ 6.7 \\
$\gamma$ (m s$^{-1}$) & --3594.6$\pm$ 1.0 \\
$e$ & 0 (adopted) ($<$0.04 at 3$\sigma$)\\ 
$M_{\rm *}$ (M$_{\rm \odot}$) & 0.58 $\pm$ 0.05\\
$R_{\rm *}$ (R$_{\rm \odot}$) & 0.598$^{+ 0.034}_{- 0.042}$\\
$\log g_{*}$ (cgs) & 4.646$^{+ 0.059}_{- 0.044}$\\
$\rho_{\rm *}$ ($\rho_{\rm \odot}$) & 2.70$^{+ 0.61}_{- 0.36}$\\
$M_{\rm P}$ (M$_{\rm Jup}$) & 1.78$\pm$ 0.10 \\
$R_{\rm P}$ (R$_{\rm Jup}$) & 0.93$^{+ 0.07}_{- 0.09}$\\
$\log g_{\rm P}$ (cgs) & 3.672$^{+ 0.081}_{- 0.059}$\\
$\rho_{\rm P}$ ($\rho_{\rm J}$) & 2.21$^{+ 0.73}_{- 0.41}$\\
$a$ (AU)  & 0.0142 $\pm$ 0.0004 \\
$T_{\rm P, A=0}$ (K) & 1370 $\pm$ 70\\
\hline 
\multicolumn{2}{l}{Errors are 1$\sigma$; Limb-darkening coefficients were:}\\
\multicolumn{2}{l}{(EulerCAM) a1 =    0.694, a2 = --0.695, a3 =  1.368, 
a4 = --0.527}\\
\multicolumn{2}{l}{(TRAPPIST) a1 = 0.748, a2 = --0.697, a3 =  
1.103, a4 = --0.438}
\end{tabular} 
\end{table}

\section{The star WASP-43}
The 15 CORALIE spectra of WASP-43 were co-added to produce a spectrum
with a typical S/N of 70:1, which we analysed using the methods
described in Gillon et\,al.\ (2009).  We used the H$\alpha$ line to
determine the effective temperature ($T_{\rm eff}$), and the Na\,{\sc
i}\,D and Mg\,{\sc i}\,b lines as diagnostics of the surface gravity
($\log g_{\rm *}$). The parameters obtained are listed in Table~2. The
elemental abundances were determined from equivalent-width
measurements of several clean and unblended lines. A value for
microturbulence ($\xi_{\rm t}$) was determined from Fe\,{\sc i} using
Magain's (1984) method. The quoted error estimates include that given
by the uncertainties in $T_{\rm eff}$, $\log g_{\rm *}$ and $\xi_{\rm t}$, as
well as the scatter due to measurement and atomic data uncertainties.

The temperature and $\log g_{\rm *}$ values, and also the {\sl
BVRIJHK\/} magnitudes collected by SIMBAD,  are consistent with a K7
main-sequence star.  The star might have
slightly above-Solar metal abundances (Table~2).  There is no
detection of lithium in the spectra, with an equivalent
width upper limit of 18m\AA, corresponding to an abundance upper limit
of log A(Li) $<$ 0.2 $\pm$ 0.3. For mid-K stars lithium is expected to
be depleted in only a few 100s Myr.  The presence of strong Ca H+K
emission indicates that WASP-43 is an active star.

The projected stellar rotation velocity ($v\,\sin I$) was determined
by fitting the profiles of several unblended Fe\,{\sc i} lines. 
Macroturbulence was assumed to
be zero, since for mid-K stars it is expected to be lower than that of
thermal broadening (Gray 2008). An
instrumental FWHM of 0.11 $\pm$ 0.01 \AA\ was estimated from the telluric
lines around 6300\AA. The best-fitting value of $v\,\sin I$ was 4.0 
$\pm$ 0.4 km\,s$^{-1}$.
For a K7V star, $V$ = 12.4 would indicate a distance of $\sim$\,80 pc.
The proper motion of 0.06$^{\prime\prime}$\,yr$^{-1}$ (Zacharias
et\,al.\ 2010) then indicates a transverse velocity of 23
km\,s$^{-1}$, which is typical of a local thin-disk star.

\subsection{Rotational modulation}
We searched the WASP photometry of WASP-43 for rotational
modulations by using a sine-wave fitting algorithm as described
by Maxted \etal\ (2011). We estimated the
significance of periodicities by subtracting the fitted transit lightcurve
and then repeatedly and randomly permuting the nights of observation.  

The 2009 WASP-South data show a modulation at a period of
15.6 $\pm$ 0.4 d with a significance of $>$99.9\%. The amplitude
is 0.006 $\pm$ 0.001 mag. The 2010 data show the same
modulation less clearly, but still with a 
significance of $>$99\%.   Most likely 15.6 d is the rotation
period of WASP-43.  Formally we cannot exclude the possibility
that the 15.6-d modulation is the beat period between the 1-d sampling
and the true period, which would then be either 0.94 or 1.07 d,
but such fast rotation would be expected to result in H$\alpha$ in
emission, whereas the CORALIE spectra show it to be in absorption,
and, further, this rotation period would be incompatible with 
the measured $v\,\sin I$ of 4.0 $\pm$ 0.4 \kmps\ unless the star were
nearly pole-on.  

Taking the rotation period as 15.6 d, and with the radius from
Table~2, gives an equatorial velocity of $v$ = 2.0 $\pm$ 0.1 \kmps.
This is significantly lower than the measured \vsini, which suggests
that there is additional broadening in the lines.  For example the two
are reconciled by increasing macroturbulence to 3 \kmps.  
The 15.6-d
rotation rate implies a gyrochronological age of 
$400^{+200}_{-100}$~Myr, using the Barnes (2007) relation. 
However, this would assume that the star has not been spun up
by tidal interactions with the planet; an alternative possibility
is that the star could be significantly older, but has been
spun up by tidal interactions that will eventually destroy the
planet (see, e.g., Brown \etal\ 2011).

\section{System parameters}
The CORALIE radial-velocity measurements were combined with the WASP,
Euler and TRAPPIST photometry in a simultaneous Markov-chain
Monte-Carlo (MCMC) analysis to find the parameters of the WASP-43
system (Table~2). For details of our methods see 
Collier Cameron \etal\ (2007b). For limb darkening we used the 
4-parameter non-linear
law of Claret (2000) with parameters at the values noted in
Table~2. 

The data are compatible with zero eccentricity (with a
3$\sigma$ limit of 0.04) and thus a circular orbit was imposed on the
solution in Table~2.  The fitted parameters were thus $T_{\rm c}$,
$P$, $\Delta F$, $T_{14}$, $b$, $K_{\rm 1}$, where $T_{\rm c}$ is the
epoch of mid-transit, $P$ is the orbital period, $\Delta F$ is the
fractional flux-deficit that would be observed during transit in the
absence of limb-darkening, $T_{14}$ is the total transit duration
(from first to fourth contact), $b$ is the impact parameter of the
planet's path across the stellar disc, and $K_{\rm 1}$ is the stellar
reflex velocity semi-amplitude.

In Fig.~3 we plot the location of WASP-43 on a modified
H--R diagram, where $T_{\rm eff}$ comes from the spectroscopy
and the stellar density from fitting the transit.  We have 
then used the density, $T_{\rm eff}$, and a metallicity of $Z = 0.017$
to fit to the evolutionary tracks from Girardi \etal\ (2000) in order 
to obtain stellar radius and mass. 
The result is a self-consistent set of parameters (Table~2; Fig.~3)
which show that WASP-43 is consistent with a main-sequence K7 star,
though we caution that stellar mass--radius calibrations are
not well constrained in this mass range.

\begin{figure}
\hspace*{-5mm}\includegraphics[width=9cm]{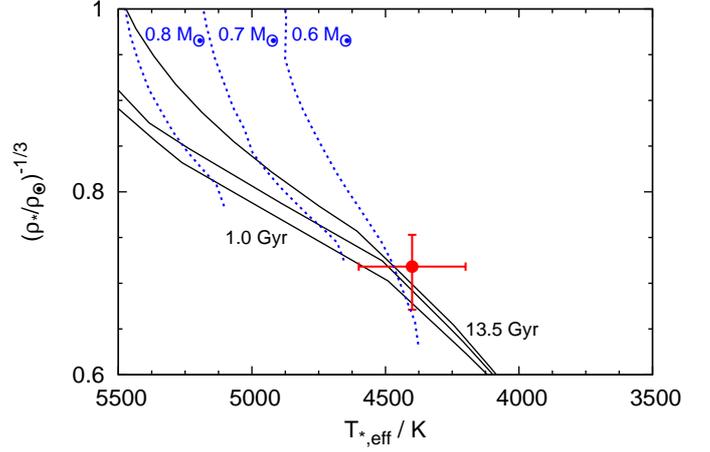}
\caption{Evolutionary tracks on a modified H--R diagram
($\rho_{*}^{-1/3}$ versus $T_{\rm eff}$).  The isochrones  
(at 1, 5 and 13.5 Gyr) and mass tracks (0.6, 0.7 \&\ 0.8 M$_{\odot}$)
are for $Z$ = 0.017 from Girardi \etal\ (2000).}
\end{figure}

\section{Discussion}
WASP-43b has an exceptionally
short orbital period of 0.81 d, second only to that of WASP-19b 
(0.79 d, Hebb \etal\ 2010) among confirmed planets. 
The host star, at K7V, has the lowest mass of any star orbited
by a hot Jupiter, and this results in the planet 
having the smallest semi-major
axis of any hot Jupiter. Among all known planets only the 
super-Earth GJ\,1214b (Charbonneau \etal\ 2009) has a value as small
(see Fig.~4). 

Hot Jupiters with such short periods are much rarer than 
those in an apparent `pile up' near 3--4 d. For example
in Hellier \etal\ (2011b) we estimated that they were
two orders of magnitude less common, and found only
because ground-based transit surveys such as WASP are most
sensitive to the shortest-period planets.
For comparison the {\sl Kepler\/} list of 1235 planet candidates
from 156,000 stars shows no systems with parameters similar 
to those of WASP-43b and WASP-19b (Borucki \etal\ 2011; Howard
\etal\ 2011).\footnote{Though there is one possibly even
more extreme system, the multi-planet candidate KOI-961, with
a possible Jupiter-sized planet orbiting a K star with 
a period of only 0.45 d (Borucki \etal\ 2011; Lissauer \etal\
2011). At magnitude 15.9 this system will be 
hard to confirm.}

Ford \&\ Rasio (2006) noted that the decline in the population
of hot Jupiters at short orbital periods occurs near 2 Roche
radii (2\,$a_{\rm R}$), which is the radius at which planets 
scattered into eccentric orbits from
much further out would tend to circularize.     The idea that
many of the hot Jupiters in the 3--4-d pileup have undergone
scattering or Kozai migration (e.g.\ Fabrycky \&\ Tremaine 2007; Nagasawa
\etal\ 2008) is supported by the finding
that many have orbits which are misalinged with the stellar
spin axes (e.g.\ Triaud \etal\ 2010; Winn \etal\ 2010), which cannot
be explained by simple  disk migration.

WASP-43b is currently at 2.1 $a_{\rm R}$ (using the definition of 
Ford \&\ Rasio 2006), and thus fits the scenario of
circularization near this radius after third-body interactions
(e.g.\ Matsumura \etal\ 2010; Guillochon \etal\ 2010; Noaz \etal\
2011).   Thus, relative to $a_{\rm R}$, WASP-43b is not 
unusually close, and the small semi-major axis results from
the unusually low mass of the host star.   The rarity of
objects like WASP-43b then indicates that hot Jupiters are
rare around low-mass stars, or that their small orbital
separations mean that their lifetimes owing to tidal decay
are short.  

Tidal theory tells us that planets as close as WASP-43b
will be spiralling inwards on a timescale
set largely by the efficiency of tidal dissipation within the
star (e.g.\ Rasio \etal\ 1996; Matsumura \etal\ 2010). 
The stellar dissipation is denoted by the quality factor,
\qps.  Thus the tidal inspiral time for WASP-43b from
its current location (e.g.\ eqn 5 of Levrard \etal\ 2009) 
is 8 Myr for \qps\ = 10$^{6}$, 80 Myr for \qps\ = 10$^{7}$,
and 800 Myr for \qps\ = 10$^{8}$.    A value of \qps\ = 10$^{6}$
has often been applied to planets (e.g.\ Levrard \etal\ 2009)
based on calibrations from binary stars.   However, such
a value would give implausibly low lifetimes for planets
such as WASP-19b (Hebb \etal\ 2010), WASP-18b (Hellier \etal\ 2009),
and now WASP-43b. 

Further, theorists have argued (e.g.\ Barker \& Ogilvie 2009; Penev \&\ 
Sasselov 2011) that values of \qps\ from binary
stars are inappropriate, since stellar-mass companions spin
up the stars to near the tidal forcing frequency, whereas
planetary-mass companions do not.  The dissipation
of tidal waves is strongly affected by resonance between
the tidal forcing and internal stellar waves (e.g.\ Sasselov
2003; Ogilvie \&\ Lin 2007; Barker \&\ Ogilvie 2009; Penev \&\ 
Sasselov 2011), and so
values of \qps\ of 10$^{8}$--10$^{9}$ might be more appropriate
in the case of planets.   Values of \qps\ $>$\,5\,$\times$\,10$^{7}$ would
result in WASP-43b having a lifetime comparable to or greater 
than the current gyrochronological age of its star. 

The main novelty contributed by WASP-43b is that, at
K7, the host star is of much later spectral type than
the other stars hosting planets with apparently short tidal lifetimes.  
For example WASP-18, which has the strongest tidal interaction
of any planet--star system, is an F6 star, while WASP-19, WASP-12 
and OGLE-TR-56 are G stars.   The tidal dissipation will
depend on the depth of the convection layer and its
interface with a radiative zone,  and thus values for 
\qps\ would be expected to depend on spectral
type (e.g.\ Barker \& Ogilvie 2009).  With a later
spectral type, and having a known rotation period,
WASP-43 will be an important test case for theoretical
\qps\ values. 

\begin{figure}
\hspace*{-5mm}\includegraphics[width=9cm]{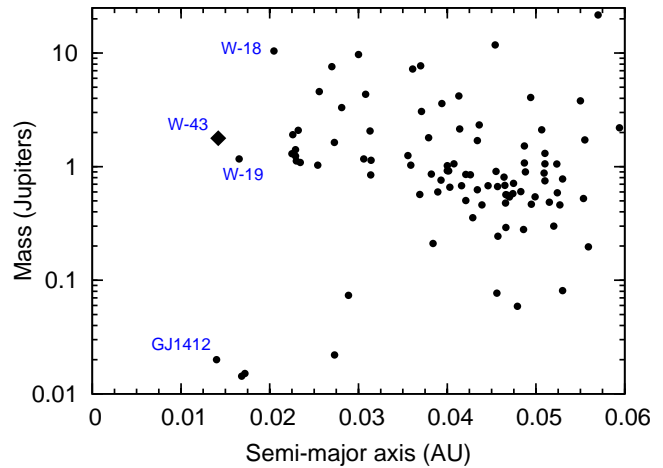}
\caption{Mass versus semi-major axis for confirmed planets, where
both are known, as compiled by Schneider (2011, as of March).
Some planets are labelled with abbreviated names.}
\end{figure}

\begin{acknowledgements}
%\section*{Acknowledgments}
WASP-South is hosted by the South African
Astronomical Observatory while SuperWASP-North is
hosted by the Instituto de Astrof\'isica de Canarias, and
we are grateful for their ongoing support and assistance. 
Funding for WASP comes from consortium universities
and from the UK's Science and Technology Facilities Council.
TRAPPIST is funded by the Belgian Fund for Scientific  
Research (Fond National de la Recherche Scientifique, FNRS) under the  
grant FRFC 2.5.594.09.F, with the participation of the Swiss National  
Science Fundation (SNF).  M. Gillon and E. Jehin are FNRS Research  
Associates.
\end{acknowledgements}


\begin{thebibliography}{}
\bibitem[Bakos et al.(2002)]{hat} Bakos G. \`{A}., L\`{a}z\`{a}r J., Papp I., S\`{a}ri P., Green E. M., 2002, \pasp, 114, 974 
\bibitem[Barnes(2007)]{2007ApJ...669.1167B} Barnes, S.A. 2007, \apj, 669, 1167
\bibitem{barker}Barker, A.J., Ogilvie, G.I. 2009, MNRAS, 395, 2268
\bibitem{borucki}Borucki, W. J. \etal\ 2011, ApJ submitted (arXiv:1102.0541) 
\bibitem{brown}Brown, D.J.A., Collier Cameron, A., Hall, C., Hebb, L., 
Smalley, B., 2011, MNRAS, in press (arXiv:1103.3599) 
\bibitem[Charbonneau et al.\ (2009)]{GJ1214b}Charbonneau, D.,  et al., 2009, Nature, 462, 891	
\bibitem[Claret (2000)]{claret}Claret, A., 2000, A\&A, 363, 1081
\bibitem[Collier Cameron et al.(2007a)]{wasp1}Collier Cameron, A., et al., 2007a, \mnras, 375, 951 
\bibitem[Collier Cameron et al.(2007b)]{mcmc}Collier Cameron, A., et al., 2007b, \mnras, 380, 1230 
\bibitem[Fabrycky \& Tremaine (2007)]{fabrycky}Fabrycky, D. \&\ Tremaine, S., 2007, ApJ, 669, 1298
\bibitem[Ford \& Rasio (2006)]{fordrasio}Ford, E. B. \&\ Rasio, F. A., 2006, 
ApJ, 638, L45
\bibitem[Gillon et al.\ (2007)]{spectral}Gillon, M., et al., 2009, A\&A, 496, 259
\bibitem{giradi}Girardi, L., Bressan, A., Bertelli, G., Chiosi, C. 2000, A\&AS,
 141, 371
\bibitem[Gray (2008)]{gray}Gray, D.F., 2008, The observation and analysis of stellar photospheres, 3rd Edition, CUP, p.~507.
\bibitem[Guillochon et al.\ (2010)]{guill}Guillochon, J., Ramirez-Ruiz, E., Lin, D. N. C., 2010,  ApJ, submitted (arXiv:1012.2382)
\bibitem[Hebb et al.\ (2010)]{hebb}Hebb, L. et al., 2010, ApJ, 708, 224
\bibitem[Hellier et al.\ (2009)]{w18}Hellier, C. et al., 2009, Nature, 460, 1098\bibitem[Hellier et al.\ (2011a)]{hohp}Hellier, C. et al., 2011a, in 
``Detection and dynamics of transiting exoplanets", eds F. Bouchy, R. D\'iaz,
 C. Moutou, EPJ Web of Conferences, Volume 11, id.01004
\bibitem{w19rm}Hellier, C.,  Anderson, D.R., Collier-Cameron, A., 
Miller, G.R.M., Queloz, D., Smalley, B., Southworth, J., Triaud, A.H.M.J.,
2011b, ApJL, 730, L31
\bibitem{howard}Howard, A. W. \etal\ 2011, ApJ submitted (arXiv:1103.2541) 
\bibitem[Levrard et al.\ (2009)]{levrard}Levrard, B. Winisdoerffer, C. \&\ Chabrier, G., 2009, ApJ, 692, L9
\bibitem{lissaur}Lissauer, J.J. \etal\ 2011, ApJ submitted (arXiv:1102.0543)
\bibitem[Magain (1984)]{magain}Magain, P., 1984, A\&A, 134, 189
\bibitem{mats}Matsumura, S., Pudritz, R.E.. Thommes, E.W. 2007, ApJ, 660, 1609
\bibitem[Matsumura et al.\ (2010)]{matsumura}Matsumura, S., Peale, S. J., Rasio, F. A., 2010, ApJ, 725, 1995
\bibitem{maxted}Maxted, P.F.L. \etal\ 2011, PASP, in press (arXiv:1012.2977) 
\bibitem[Nagasawa et al.\ (2008)]{nagasawa}Nagasawa, M., Ida, S., Bessho, T., 2008, ApJ, 678, 498
\bibitem{naoz}Naoz, S., Farr, W. M., Lithwick, Y., Rasio, F. A., Teyssandier, J., 2011, Nature, in press (arXiv:1011.2501)
\bibitem[Penev \&\ Sasselov (2011)]{penev}Penev, K. \& Sasselov, D., 2011, ApJ, in press (arXiv:1102.3187) 
\bibitem[Pollacco et al.\ (2006)]{sw}Pollacco, D., et al., 2006, \pasp, 118, 1407 
\bibitem[Pollacco et al.\ (2007)]{wasp3}Pollacco, D., et al., 2008, \mnras, 385, 1576
\bibitem[Queloz et al.\ (2001)]{bisector}Queloz, D., et al., 2001, \aap, 379, 279
\bibitem[Rasio et al.\ (1996)]{rasio}Rasio, F. A., Tout, C. A., Lubow, S. H., 
     Livio, M., 1996,  ApJ, 470, 1187
\bibitem{sasselov}Sasselov, D.D. 2003, ApJ, 596, 1327 
\bibitem[Schneider (2011)]{schneider}Schneider, J., 2011, http://exoplanet.eu/catalog-transit.php 
\bibitem[Triaud et al.\ (2010)]{triaud10}Triaud, A. H. M. J.,  2010, A\&A, 524, 25
\bibitem[Winn et al.\ (2010)]{winn10}Winn, J. N., Fabrycky, D., Albrecht, S., 
Johnson, J. A.,  2010, ApJ, 718, L145
\bibitem{zacharias}Zacharias, N. \etal\ 2010, AJ, 139, 2184

\end{thebibliography}
\end{document}